\documentclass[pra,twocolumn,floatfix]{revtex4}
\usepackage{amsmath,amssymb,amsfonts,bbm,graphicx,hyperref,times,psfrag}
\usepackage{amsthm}
\usepackage{amsfonts}
\usepackage{microtype}
\usepackage{float}
\usepackage{mathrsfs}
\usepackage{placeins}
\usepackage{bbm}
\usepackage{color}
\usepackage{epstopdf}

\newcommand{\be}{\begin{equation}}
\newcommand{\ee}{\end{equation}}

\newcommand{\daga}{^{\dagger}}
\newcommand{\bsigma}{\boldsymbol \sigma}
\newcommand{\bbeta}{\boldsymbol \beta}

\def\sts{\!\hbox{\tiny STS}\,}
\def\sv{\!\hbox{\tiny SV}\,}
\def\es{\hbox{\tiny E}\,}
\def\tc{{\hbox{\tiny C}\,}}
\def\tl{{\hbox{\tiny L}\,}}
\def\tq{{\hbox{\tiny Q}\,}}
\def\ta{\hbox{\scriptsize A}}
\def\tb{\hbox{\scriptsize B}}

\begin{document}
\title{Entanglement as a resource for discrimination of classical environments}
\author{Jacopo Trapani}\email{jacopo.trapani@unimi.it}
\affiliation{Quantum Technology Lab, Dipartimento di Fisica, 
Universit\`a degli Studi di Milano, I-20133 Milano, Italy}
\author{Matteo G.~A.~Paris}\email{matteo.paris@fisica.unimi.it}
\affiliation{Quantum Technology Lab, Dipartimento di Fisica, 
Universit\`a degli Studi di Milano, I-20133 Milano, Italy}
\affiliation{INFN, Sezione di Milano, I-20133 Milano, Italy}
\date{\today}
\begin{abstract}
We address extended systems interacting with classical fluctuating
environments and analyze the use of quantum probes to discriminate
{\em local noise}, described by independent fluctuating 
fields, from {\em common noise}, corresponding to the interaction 
with a common one. In particular, we consider a bipartite system 
made of two non interacting harmonic oscillators and assess 
discrimination strategies based on homodyne detection, 
comparing their performances with the ultimate bounds on the error
probabilities of quantum-limited measurements.
We analyze in details the use of Gaussian probes, with 
emphasis on experimentally friendly signals.
Our results show that a joint measurement of the position-quadrature
on the two oscillators outperforms any other homodyne-based 
scheme for any input Gaussian state. 
\end{abstract}
\maketitle
\section{Introduction}\label{s:intro}
The effects of the interaction of quantum systems with their
environments have been widely studied in the last decades. In general,
environment-induced decoherence \cite{wei01,bre01} is detrimental for
the quantum features of a localized system: loss of nonclassicality
\cite{har01,zur01, zur02} or disentanglement may arise asymptotically or
after a finite interaction time \cite{ser1,ser2,ser23,ebe10}. On the other
hand, extended systems experience more complex decoherence phenomena:
the subparts of a system may interact with indepenent environments or,
more interestingly, with a common one, corresponding to collective 
decoherence or dissipation, which may result in preservation of 
quantum coherence as well as 
preservation and creation of entanglement \cite{bra01,zha01,flo01,
pra01,con01,paz01} or superradiance \cite{bro01,gro01,pal01,
riv01,zan01,dua01,kwi01,dic01}.
\par
Decoherence and dissipation into a common bath 
may arise spontaneously in some structured environments, 
but it may also be engineered \cite{bar01, ver01} to 
achieve specific goals. In both cases, the common decoherence 
mechanism may mingle or even being overthrown by local 
processes, leading to undesidered loss of quantum features. 
The discrimination between the presence of local or common 
environments is thus a relevant tool to fight 
decoherence and preserve quantum coherence.
\par
Describing the interaction with an external environment in a full
quantum picture may be challenging.  On the other hand, in many
situations the action of the environment on a quantum system may be
represented as an external random force on the system itself.  Such
random forces are described in terms of classical stochastic fields
(CSF)\cite{gar01}.  
As a matter of fact, the description of a quantum
environment in terms of CSFs is often very accurate 
in capturing the quantum features of the dynamics. 
Besides, many system-environment interactions
have a classical equivalent description \cite{hel01, hel02, cro01, wit01,
str02, sto01} and there are situations where the 
environment can be effectively simulated classically 
\cite{tur01}. Finally, we mention that in several situations 
of experimental
interest \cite{ast01, gal01, abe01, str01} quantum systems 
interact with inherently classical Gaussian noise.
\par
In this framework, the main goal of this paper is to 
design a successful strategy to discriminate which 
kind of interaction, either local or common, occurs
when an extended quantum probe interacts with a 
classical fluctuating environment.
This is a channel discrimination problem, which we 
address upon considering a quantum probe interacting with either
a local or a common bath, and then solving the corresponding
state discrimination problem.
In particular, in order to assess the role of 
entanglement with nearly analytic results, we consider 
a bipartite system made of
two non interacting harmonic oscillators. 
The local noise scenario is
described by the interaction of each oscillator 
with independent CSFs whereas common noise 
is described as the coupling between the two oscillators 
with the same CSF.  The dynamics of this model has been 
analyzed recently \cite{tra01} revealing the
existence of a rich phenomenology, which turns out to be
a resource for discrimination purposes.
\par
The lowest probability of error achievable in a quantum discrimination
problem is known as Helstron bound \cite{hel03}. In several situations,
such bound can be approximated by the Quantum Chernoff Bound (QCB)
\cite{aud02,pir01}, originally derived in the setting of
asymptotically many copies.  Despite being less precise, the QCB turns
out to be more versatile: it is easier to evaluate and can be used as
distinguishability measure between qubits and single-mode Gaussian
states \cite{cals01,aud01}, and for these reasons it constitutes a
benchmark in quantum discrimination.
On the other hand, the Helstrom bound 
may be challenging from the experimental point of view 
and a question arises about 
the performances achievable using feasbile
measurements and realistic probe preparations.
\par
In this paper, we analyze in details discrimination strategies based
on homodyne detection, which has already proven to be useful in
discrimination of quantum states \cite{witt01} or binary communication
schemes \cite{oliv01}.  Also, we analyze in details the performances 
of Gaussian states used as probe preparation,  
including many lab-friendly input signals.
\par
Our results show that a joint measurement of the position
on the two oscillators outperforms any other homodyne-based 
scheme, whatever input Gaussian state of the probe is employed. 
In terms of error probability, a discrimination scheme based 
on homodyne detection easily outperforms the QCB using (entangled) 
squeezed thermal states as input preparation.
\par
The paper is organized as follows. In Sec. \ref{s:inter}
we introduce the interaction model, we discuss the dynamics
of the system in both local and common scenarios and
describe the classes of Gaussian states we use later on in the paper.
in Sec. \ref{sec:disc} we introduce the necessary tools
of discrimination theory. In Sec. \ref{s:double} we 
build step by step the discrimination strategy and check its
performance with some lab-friendly Gaussian input states.
Finally, in Sec, \ref{s:random}, we optimize the discrimination strategy
looking for the optimal Gaussian input state.
Section \ref{s:out} then closes the paper 
with some concluding remarks.
\section{The Interaction Model}
\label{s:inter}
We consider two non-interacting harmonic quantum oscillators with 
natural frequencies $\omega_1$ and $\omega_2$ and describe the dynamics 
of this system in two different regimes: in the first one each 
oscillator is coupled to one of two independent non-interacting 
stochastic fields: this scenario is dubbed as {\em local noise} case. 
In the second regime, the oscillators are coupled to the same classical 
stochastic field, so we dub this case as {\em common noise}.
In both case, the Hamiltonian $H$ is composed by a free and 
an interaction term. 
The free Hamiltonian $H_0$ is given by
\begin{align}
H_0 &= \hbar \sum_{j=1}^2  \omega_j  a_j \daga a_j.
\end{align}
in both regimes, whereas 
the interaction term $H_I$ differs. In the following, we 
introduce the local and the common interaction Hamiltonians.
\subsection{Local Interaction}
The interaction Hamiltonian $H_{\tl}$ in the local model reads
\begin{align}
H_{\tl} (t)& = \sum_{j=1}^2 a_j
\bar{C_j}(t) e^{i \delta_j t}+ a\daga_j C_j (t) e^{-i \delta_j t}
\end{align}
where the annihilation operators $a_1,a_2$ represent the oscillators,
each one coupled to a different local stochastic field $C_j (t)$ with $j =
1,2$, and $\delta_j = \omega_j-\omega$ is the detuning between the
carrier frequency of the field and the natural frequency of the $j$-th
oscillator. 
Throughout the paper, we will
consider the  Hamiltonian rescaled in units of energy 
$\hbar \omega_0$ (for a reason to be pointed out later).  Under
this condition, the stochastic fields $C_1 (t), C_2(t)$, their central
frequency $\omega$, the interaction time $t$, and the detunings all 
become dimensionless quantities.
\par
The presence of fluctuating stochastic fields leads to 
an explicitly time-dependent Hamiltonian, whose corresponding 
evolution operator is given by
\begin{align}
\label{eq:evolutionoperator}
U_{\tl}(t) = {\cal T} \exp\left\{ - 
i \int_0^t \!\! ds\, H_{\tl}(s) \right\}\,,
\end{align}
where ${\cal T}$ is the time ordering. 
The evolved density operator is formally given by
\be
\rho_{\tl}(t) = U_{\tl}(t) \rho(0) U\daga_{\tl}(t).
\ee
The explicit form of the density operator 
can be found following the very same steps described in \cite{tra01}.
The evolution of the density operator of the system then reads
\begin{align}
\rho_L (t) &=  
\left[D(\phi_a, \phi_b)\rho_0 D\daga (\phi_a, \phi_b) \right]_{F}  
\end{align}
where $D_j(\alpha) = \exp(\alpha a_j\daga - \alpha^{*}a_j)$ is the
displacement operator, $D(\alpha_1, \alpha_2) = D(\boldsymbol{\alpha}) =
D_1(\alpha_1) D_2 (\alpha_2)$ and $\left[\ldots \right]_{F}$ is the
average over the realizations of the stochastic fields.  
\par 
In the local scenario, we assume each CSF 
$$C_j(t) = C_j^{(x)}(t) + i C_j^{(y)}(t),$$
described as a Gaussian stochastic process
with zero mean 
$ [C_j^{(x)}(t)]_{F} =  [C_j^{(y)}(t)]_{F}=0$ and 
autocorrelation matrix given by
\begin{align}
\left[C_j^{(x)}(t_1)C_k^{(x)}(t_2) \right]_F &= 
\left[C_j^{(y)}(t_1)C_k^{(y)}(t_2) \right]_F \notag 
\\ & = \delta_{jk} K(t_1,t_2) \\
\left[C_j^{(x)}(t_1)C_k^{(y)}(t_2) \right]_F 
&= \left[C_j^{(y)}(t_1)C_k^{(x)}(t_2) \right]_F =0 
\end{align}
where we introduced the kernel autocorrelation 
function $K(t_1,t_2)$.
Upon performing the stochatic average, one finally 
recovers a Gaussian map describing the evolution of the 
state of the system under the assumption of local interaction
\be
\label{locdyn}
\rho_{\tl}(t) = \mathcal{G}_{\tl}[\rho(0)] = 
\int \frac{d^4 \boldsymbol{\zeta}}{\pi^2}
g_{\tl}(\boldsymbol{\zeta}) D(\boldsymbol{\zeta})
\rho(0) D\daga(\boldsymbol{\zeta})
\ee
where $g_{\tl}(\boldsymbol{\zeta})$
is the Gaussian function
\be
g_{\tl}(\boldsymbol{\zeta})= \frac{\exp(-\frac12 \, \boldsymbol{\zeta}
\cdot \boldsymbol{\Omega} \cdot \boldsymbol{\sigma}_{\tl}^{-1}\cdot
\boldsymbol{\Omega}^T \cdot \boldsymbol{\zeta}^T)}{\sqrt{\mbox{det}[
\boldsymbol{\sigma}_{\tl}]}} \ee
$\boldsymbol{\sigma}_{\tl}$ and the symplectic matrix 
$\boldsymbol{\Omega}$ being given by
\be 
\boldsymbol{\Omega} = \left(
\begin{array}{cc}
0 & 1 \vspace{0.1 cm}
\\
-1 & 0 
\end{array}
\right)
\quad
\boldsymbol{\sigma}_{\tl} = \left(
\begin{array}{cc}
\beta_1 (t) \mathbb{I}_2 & 0 \vspace{0.1 cm} \\
0 & \beta_2(t) \mathbb{I}_2
\end{array} \right).
\ee
The matrix $\boldsymbol{\sigma}_{\tl}$  is the covariance of the noise 
function $g_{\tl}(\boldsymbol{\sigma})$
and its matrix elements are given by 
\begin{align}
&\beta_j (t,t_0)= \int_{t_0}^t \int_{t_0}^t d s_1 d s_2 
\cos[\delta_j(s_1 - s_2)] K(s_1,s_2).
\end{align}
\subsection{Common Interaction}
The interaction Hamiltonian $H_{\tc}$ for the common noise case
reads as follows 
\begin{align}
H_{\tc} (t)& = \sum_{j=1}^2 a_j  e^{i \delta_j t}\bar{C}(t) + a\daga_j  e^{-i \delta_j t}C (t)
\end{align}
where each oscillator, represented by the
annihilation operators $a_1,a_2$, is coupled to a common stochastic
field $C(t)$ which is described as a Gaussian stochastic 
process with zero mean $[C^{(x)}]_F = [C^{(y)}]_F =0 $ and  the very same
autocorrelation matrix of the local scenario.
\par
Along the same lines of the local interaction model derivation,
we find the Gaussian map that describes the evolution of 
the state of the system
\be
\label{comdyn}
\rho_{\tc}(t) = \mathcal{G}_{\tc}[\rho(0)] = \int 
\frac{d^4 \boldsymbol{\zeta}}{\pi^2}g_{\tc}
(\boldsymbol{\zeta}) D(\boldsymbol{\zeta})\rho(0) 
D\daga(\boldsymbol{\zeta})
\ee
where $g_{\tc}(\boldsymbol{\zeta})$ 
the Gaussian function
\be
g_{\tc}(\boldsymbol{\zeta})= \frac{\exp(-\frac12 \, \boldsymbol{\zeta}
\cdot  \boldsymbol{\Omega} \cdot \boldsymbol{\sigma}_{\tc}^{-1}\cdot
\boldsymbol{\Omega}^T \cdot \boldsymbol{\zeta}^T)}{\sqrt{\mbox{det}[
\boldsymbol{\sigma}_{\tc}]}} 
\ee
 $\boldsymbol{\sigma}_{\tc}$ being its covariance matrix, 
given by
\begin{align}
&\boldsymbol{\sigma}_{\tc} = \left(
\begin{array}{cc}
\beta_1 (t) \mathbb{I}_2 & \boldsymbol R \vspace{0.1 cm} \\
\boldsymbol R & \beta_2(t) \mathbb{I}_2
\end{array} \right)
\\
&\boldsymbol R = \left(
\begin{array}{cc}
\beta_{\tc} (t)  & \gamma_{\tc} (t) \vspace{0.1 cm} \\
\gamma_{\tc} (t) & \beta_{\tc}(t) 
\end{array} \right)
\end{align}
with the matrix elements given by
\begin{align}
\beta_c (t,t_0)=
 \int_{t_0}^t \int_{t_0}^t & d s_1 d s_2 
\cos[(\delta_1 s_1 - \delta_2 s_2)]K(s_1,s_2) \nonumber
\end{align}
\begin{align}
\gamma_c (t,t_0)=
 \int_{t_0}^t \int_{t_0}^t & d s_1 d s_2 
\sin[(\delta_1 s_1 - \delta_2 s_2)] K(s_1,s_2). \nonumber
\end{align}
\subsection{Dynamics in the local and the common noise scenarios}
The dynamical maps described by Eqs. (\ref{locdyn}) and (\ref{comdyn}) 
correspond to Gaussian channels,which represent the short times
solution of Markovian (dissipative) Master equations in the limit of
high-temperature environment. In the following, this link will be
exploited to analyze the limiting behaviour of the two-mode dynamics.
\par
In order to get quantitative results, we assume that fluctuations 
in the environment are described by Ornstein-Uhlenbeck Gaussian 
processes, characterized by a Lorentzian spectrum and
a kernel autocorrelation function 
$$K(t_1,t_2) = \frac{1}{2}\lambda t_{\es}^{-1} 
\exp({|t_1 -t_2|/t_{\es}})\,,$$
where $\lambda$ is a coupling constant and $t_{\es}$ is 
the correlation time of the environment. 
We also assume the oscillators are both resonant with 
the central frequency of the stochastic field
$(\omega_1 = \omega_2 = \omega)$, i.e. that both 
detunings from the central frequency of the 
classical stochastic field are vanishing 
$$\delta_1 = \delta_2 = \delta = 1-\frac{\omega}{\omega_0} = 0\,.$$
This assumption leads to a simpler expression of the state dynamics:
in the local scenario, it leads to
$\beta_1 (t) = \beta_2 (t) = \beta(t)$ and, in turn, to 
\begin{flalign}
\label{eq:symplocdyn}
\rho_{\tl} (t) &= \mathcal{E}_{\tl}[\rho(0)](t) 
\nonumber \\ &=  \int \frac {d^2 \zeta_1 }{\pi \beta(t) } 
\int \frac {d^2 \zeta_2 }{\pi \beta(t) } 
\exp \left( - \frac{|\zeta_1|^2+|\zeta_2|^2}{\beta(t)}\right) 
\nonumber \\ &\quad \times
D(\zeta_1) \otimes D(\zeta_2) \rho(0) D\daga(\zeta_1) \otimes
D\daga(\zeta_2).
\end{flalign}
where $\beta(\Delta t = t- t_0) = \beta(t,t_0)$ is given by
\begin{align}
\beta(t) =& \, \lambda
(t-1 + e^{-t}).
\end{align}
and where $\beta(t)$ has been rescaled in units 
of $t_{\es}$, i.e. $\lambda \rightarrow \lambda\, t_{\es}$,
$t \rightarrow t/t_{\es}$.
\par
In the common noise case, the condition of resonant oscillators 
implies $\beta_1(t) = \beta_2(t) = \beta_c(t) = \beta(t)$ and
$\gamma_c (t) =0$, leading to simplified matrices 
$\boldsymbol R$ and $\boldsymbol \sigma_{\tc}$ given by
\be
\boldsymbol R = \left(
\begin{array}{cc}
\beta (t)  & 0\vspace{0.1 cm} \\
0 & \beta (t) 
\end{array} \right)
\quad
\boldsymbol{\sigma}_{\tc} = \left(
\begin{array}{cc}
\beta (t) \mathbb{I}_2 & \boldsymbol R \vspace{0.1 cm} \\
\boldsymbol R & \beta(t) \mathbb{I}_2
\end{array} \right)
\ee
corresponding to the Gaussian channel
\begin{align}
\label{eq:sympcomdyn}
\rho(t) &= \mathcal{E}_{\tc}[\rho(0)](t) =   
\nonumber \\ &=  \int \frac {d^2 \zeta }{\pi \beta(t) } 
\exp \left( - \frac{|\zeta|^2}{\beta(t)}\right) \nonumber \\ &\quad \times
D(\zeta) \otimes D(\zeta) \rho(0) D\daga(\zeta) \otimes D\daga(\zeta).
\end{align}
Both the local and the common interaction models 
correspond to Gaussian channels, i.e. they map 
Gaussian states into Gaussian states, preserving the 
Gaussian character at any time.
\subsection{Input states}
Before introducing the necessary tools for quantum 
state discrimination, we briefly discuss what kind 
of input states we are about to consider. In general, 
the optimization of a channel discrimination protocol 
involves the optimization over the possible input 
states. For continuous variable systems focussing attention
on Gaussian states is a convenient choice for at least
two reasons.  On the one hand, the evaluation of commonly 
used figures of merit
as entanglement or purity comes at ease. On the other hand, as the
dynamics is described by Gaussian channels, the dynamics
may be evaluated analytically in the covariance matrices
formalism. In fact, at any time $t$ the state is Gaussian and 
it is fully described by its
the covariance matrices
$\bbeta_{\tl} (t) $ and $\bbeta_{\tc}(t)$ for the local
and common scenario,
\begin{align}
\bbeta_{\tl} (t) &= \bsigma_{0} + 2 \, \bsigma_{\tl} (t) \label{evsl} \\
\bbeta_{\tc} (t) &= \bsigma_{0} + 2 \, \bsigma_{\tc} (t) \label{evsc}
\end{align}
where $ \bsigma_{0}$ denotess the covariance matrix of a
generic Gaussian input state.
The most generic two-mode Gaussian state is described by the covariance
matrix $\bsigma_{g}$, but every generic $\bsigma_{g}$ can be recast by
local operations in a simpler form $\bsigma_{i}$ called standard form,
\be \bsigma_{g} = \left(
\begin{array}{cc}
\boldsymbol A & \boldsymbol C  \\
\boldsymbol C^T & \boldsymbol B  
\end{array}
\right)
\quad
\bsigma_{i} = \left(
\begin{array}{cccc}
a & 0 & c &0 \\
0 & a & 0 &d \\
c & 0 & b & 0 \\
0 & d & 0 & b
\end{array}
\right)
\ee
where $\boldsymbol A, \boldsymbol B, \boldsymbol C$ are $2 \times 2$
matrices, $\bsigma_{g}$ (and $\bsigma_i$ as well) satisfies the
condition $\bsigma_{g}+ \frac{i}{2}\boldsymbol \Omega \geq 0$, with
$\boldsymbol \Omega = \boldsymbol \omega \otimes \boldsymbol \omega$ and
$\boldsymbol \omega = \{\{0, 1\}, \{ -1, 0 \} \}$.  Among 
Gaussian states, we focus attention on
three relevant classes, squeezed thermal states (STSs), states obtained
as a linear mixing of a single-mode Gaussian state with the vacuum (SVs)
and standard form SVs, i.e. SV states recast in standard form by local
operations. These classes of states may be generated by current quantum
optical technology and thus represent good candidates for the 
experimental implementations of discrimination protocols.
STSs and SVs are described by the 
covariance matrices $\boldsymbol \sigma_{\sts}$
and $\boldsymbol \sigma_{\sv}$, respectively
\be
\label{sqthermalstate}
\boldsymbol \sigma_{\sts} = \frac12 \left(
\begin{array}{cccc}
a & 0 & c &0 \\
0 & a & 0 &-c \\
c & 0 & b & 0 \\
0 & -c & 0 & b
\end{array} \right)
\quad
\boldsymbol \sigma_{\sv} = \frac14 \left(
\begin{array}{cccc}
m & 0 & s_1 &0 \\
0 & n & 0 &s_2 \\
s_1 & 0 & m & 0 \\
0 & s_2 & 0 & n
\end{array} \right).
\ee
The covariance matrix $\boldsymbol \sigma_{\sts}$ corresponds to a
density operator of the form 
\be
\rho_{\sts} = S_2 (r) (\nu_1 \otimes \nu_2) S_2 (r)\daga
\ee 
where $S_2 (r) = \exp \{ r (a_1\daga a_2\daga -a_1 a_2) \}$ is the
two-mode squeezing operator and $\nu_j$ is a single-mode thermal state 
\be
\nu_j = \frac{1}{\bar n_j}\sum_m \left(\frac{\bar n_j}{\bar n_j +1}\right)^m |m \rangle \langle m|.
\ee
The physical state depends on three real parameters: the squeezing
parameter $r$ and the two numbers $\bar n_1,\bar n_2$, which are related
to the parameters $a,b,c$ of eq. (\ref{sqthermalstate}) by the relations
\begin{align} 
a &= \cosh (2r) + 2 \bar n_1 \cosh^2 r + 2 \bar n_2 \sinh^2 r \nonumber \\
b &= \cosh (2r) + 2 \bar n_1 \sinh^2 r + 2 \bar n_2 \cosh^2 r \nonumber \\
c &= (1+ \bar n_1 + \bar n_2 ) \sinh (2 r).
\end{align}
In particular, we focus on symmetrical thermal states $\tilde{n}_1= \tilde{n}_2 = \tilde{n}$,
that can be re-parametrized setting
$\epsilon = 2( \bar n + n_s +2 \bar n\, n_s
)$, with $n_s = \sinh^2 r$, and a normalized squeezing parameter $\gamma
\in [0,1]$, such that
\begin{align}
n_s &= \gamma \epsilon \nonumber 
\quad \quad \bar n = \frac{(1-\gamma) \epsilon}{1+2 \gamma \epsilon}.
\end{align}
\par
The covariance matrix $\boldsymbol \sigma_{\sv}$ corresponds to a
density operator of the form 
\be
\rho_{\,\sv} = R\left(\frac{\pi}{4}\right) \Big( S(r)\nu S\daga (r) \otimes |0 \rangle \langle 0| \Big) R\daga \left(\frac{\pi}{4}\right)
\ee 
where $S (r) = \exp \{ r (a_1\daga -a_1 \daga) \}$ is the single-mode
squeezing operator and $R(\theta) = \exp \{ \theta(a_1 a_2\daga +
a_1\daga a_2 \}$ is the rotation operator corresponding to a
beam-splitter mixing.  The physical state depends on two real
parameters: the squeezing parameter $r$ and the number $\bar n$, which
are related to the parameters $m,n,s_1,s_2$ of eq.
(\ref{sqthermalstate}) by the relations \begin{align} 
m &= e^{2r}(1+2n)+1, \nonumber \\
n &=  e^{-2r}(1+2n)+1, \nonumber \\
s_1 &=  e^{2r}(1+2n)-1, \nonumber \\
s_2&=  e^{-2r}(1+2n)-1.
\end{align}
\par
Notice that only STSs already possess a covariance matrix
in standard form. However, simply applying local squeezing to both modes
the standard form of $\boldsymbol \sigma_{\sv}$ can be found. Of course,
locally squeezing the modes dramatically changes the energy of the Gaussian
state but leaves quantities such purity and entanglement unmodified.
We will refer to standard form single-vacuum states as SSVs.
\section{Quantum State Discrimination}
\label{sec:disc}
In this section, we briefly summarize the basic concepts of 
quantum state discrimination and introduce the tools required 
to implement a discrimination strategy.  The purpose of state 
discrimination is to distinguish, by looking at the outcome of a
measurement performed on the system, between two possible 
hypothesis on the preparation of the system itself.  
In our case, we assume to prepare
the bipartite system in a given Gaussian state and aim to distinguish 
which kind of noise, local or common, affected the system. 
This is done by a discrimination scheme applied to the output states
of the Gaussian maps (\ref{locdyn}) and (\ref{comdyn}).  
Since the two outputs are not orthogonal for any given
input, perfect discrimination is impossible and a probability
of error appears. Optimal discrimination schemes are those
minimizing the probability of error upon a suitable choice 
of both the input state and the output measurement.
The minimum achievable probability of error, given a pair of 
output states, may evaluated from the density operators of the
two state,  and it is usually referred to as the Helstrom Bound.
\par
We suppose to have a quantum system that may be prepared in
two possibile states, corresponding to the two hypotheses 
$H_{\ta}$ and $H_{\tb}$,
\begin{align}
H_{\ta} : \rho \rightarrow \rho_{\ta}
\quad
H_{\tb} : \rho \rightarrow \rho_{\tb}
\end{align}
The second step is to choose a discrimination strategy, i.e. 
one measures the system and decides among the two hypothesis 
$H\ta$ or $H\tb$.  To this purpose, one chooses a two-value 
positive-operator-valued measure (POVM) \{$E_{\ta}, E_{\tb}$\} 
with $E_{\ta} + E_{\tb} =
\mathbb{I}$ and $E_{\ta}, E_{\tb} \geq 0$.  Once the measurement is
performed, the observer infers the state of the system with an error
probability $P_e$ given by
\begin{flalign}
\label{proberr}
P_e = &\frac12 \mbox{Tr}[\rho_{\ta} E_{\tb}] 
+ \frac12\mbox{Tr}[\rho_{\tb} E_{\ta}] \\ \nonumber
 = & \frac12 (1 - \mbox{Tr}[E_{\tb} \Lambda]),
\end{flalign}
where $\Lambda$ is the Helstrom matrix,
\be
\Lambda = \rho_{\tb} - \rho_{\ta}.
\ee
The error probability is minimized for a POVM  such that
$\mbox{Tr}[E_{\tb}\Lambda] =  \frac12 \mbox{Tr}|\Lambda|$
and the minimum is given by
\be
P_e = \frac12[1- T(\rho_{\ta}, \rho_{\tb})],
\ee where
\be
T(\rho, \sigma) = \frac12 \mbox{Tr}|\rho-\sigma|
\ee
is the trace distance. $P_e$ is known as Helstrom Bound and represents
the ultimate error probability that can be ideally achieved.
Unfortunately, evaluating the Helstrom Bound for continuous variable
systems is a challenging task, as it requires performing a trace
operation on infinite matrices.  Nevertheless, some lower and upper
bounds can be found by means of the Uhlmann fidelity function
\be
\label{fidelity}
\mathcal{F}(\rho_{\ta},\rho_{\tb}) = 
\Big[ \mbox{Tr}\sqrt{\sqrt{\rho_{\ta}} \rho_{\tb} \sqrt{\rho_{\ta}}}\, \Big].
\ee 
In fact, we have \cite{pir02}
\be
\label{bounds}
\mathcal F_m \equiv
\frac{1-\sqrt{1-\mathcal{F} (\rho_{\ta},\rho_{\tb})}}{2} \leq P_e \leq
\frac{\sqrt{\mathcal{F}(\rho_{\ta},\rho_{\tb})}}{2}
\equiv \mathcal F_M.  
\ee
Another tighter upper-bound for the Helstrom Bound is given by the
quantum Chernoff bound (QCB) $Q$, \be
Q= \inf_{0 \leq s \leq 1} \mbox{Tr} [\rho^s_{\ta} \rho^{1-s}_{\tb}].
\ee
Even though the QCB does not possess any natural operational meaning,
i.e.  it cannot be directly related to a measurement process, it becomes
a powerful tool in discrimination protocols featuring multicopy states
and it is generally pretty easy to evaluate for continuous variable
systems.  The QCB can be related to the Uhlmann fidelity function and,
by means of the QCB, Eq. \ref{bounds} can be upgraded to
\be
\mathcal F_m \leq P_e \leq \frac{Q}{2} \leq \mathcal F_M,
\ee
where $\mathcal F_m$ and $\mathcal F_M$ are the 
lower and upper fidelity bounds, respectively.
The explicit formulas for the QCB and the fidelity for Gaussian states
are cumbersome and won't be reported here.
\par
The Helstrom bound represents the smallest error probability that can be
ideally achieved in state discrimination.  However, even when
evaluating the Helstrom Bound is possible, it usually corresponds
to a POVM which is difficult to implement. In the following, we 
devote attention to feasible measurements and evaluate their
performances in the discrimination of local and common noise, comparing
the error probability with the ultimate bounds discussed in this
Section.
\section{Double Homodyne Measurement}
\label{s:double}
\begin{figure}[h!]
\includegraphics[width=\columnwidth]{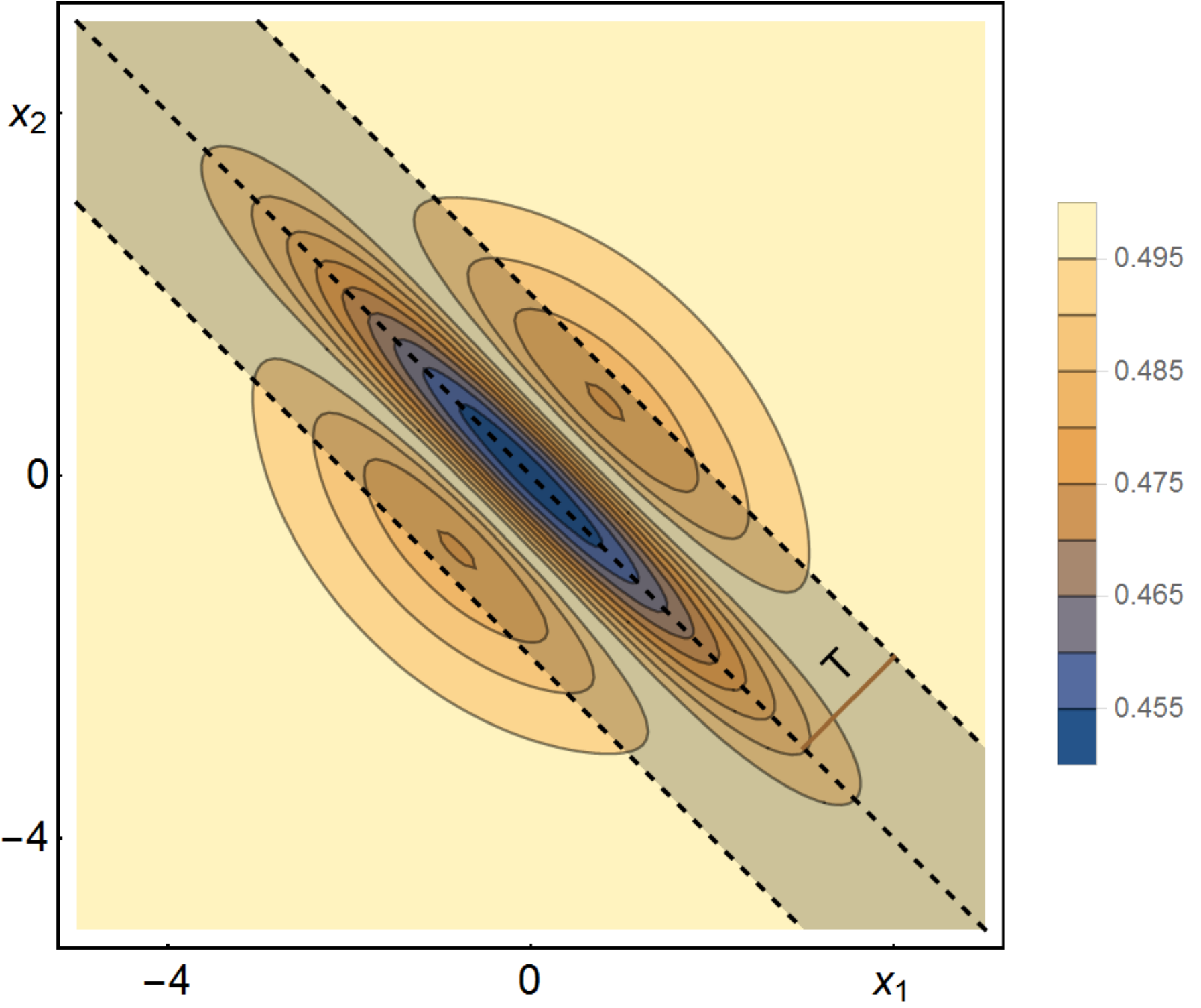}
\caption{\label{wigner} (Color Online) Contourplot of $P_\tq (x_1,x_2)$
for a STS. The dark region between the two dashed parallel lines 
represents a choice of ${\mathcal D}_c$, the region of outputs 
associated to the inference of common noise.  States and channel 
parameters are set as
follows: $\epsilon =1, \gamma = 0.7,
\lambda_1 = \lambda_2 = \lambda =1,  t =1$.} \end{figure}
\begin{figure*}[t!]
\centering
\includegraphics[width =\textwidth]{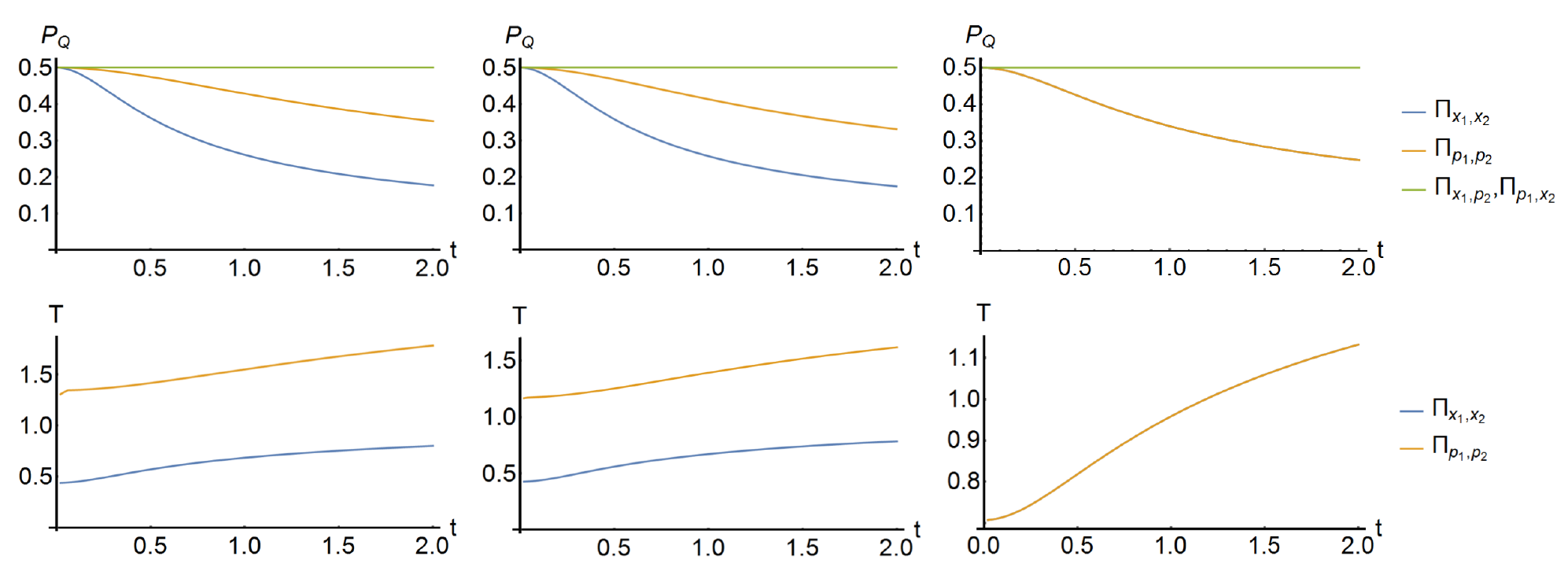}
\caption{\label{fig1} (Color Online) Upper panels: Error probability
$P_\tq$ for different POVMs for (left) STS ($\epsilon =1, \gamma =0.7$),
(center) SV ($n=1, r=0.7$) and (right) standard SV ($n=1, r=0.7$). The
POVM $\Pi (x_1,x_2)$(blue lower line) is always the most efficient.
The POVMs $\Pi (x_1,p_2)$ and $\Pi (p_1,x_2)$
(upper red and green) yield the same error probability $P_\tq = \frac12$ 
independently on the input state and are useless for discrimination 
purposes. Lower panels:
Optimal half-width $T$ as a function of time for POVMs $\Pi (x_1,x_2)$
and $\Pi (p_1,p_2)$.  We set
$\lambda_1 = \lambda_2 = \lambda =1$.} 
\end{figure*}
In section \ref{s:inter}, we have analyzed the dynamics 
in the presence of either local or common noise.
The two dynamical maps are different and, in particular,
correlations between the two oscillators appear exclusively 
in the common noise scenario, as it is apparent
from the presence of off-diagonal terms in the common noise
matrix. 
As a matter of fact, the correlation terms in the noise 
matrix corresponds the  variances  $var(X_1,X_2)$ and $var(P_1,P_2)$,
where $X_j = \frac{1}{\sqrt2} (a_j+a\daga_j), P_j = \frac{1}{i \sqrt2}
(a_j-a\daga_j) $ are the quadrature operators of the two
oscillatos. This 
argument suggests that joint homodyne detection of the quadratures
of the two modes may be a suitable building block to discriminate 
the two possible environmental scenarios. 
In the following, we are about to consider
the measurement of all possible combinations of quadratures,
$(X_1,X_2)$, $(P_1,P_2)$, $(X_1,P_2)$ and $(P_1,X_2)$ and 
denote the corresponding POVMs as
$\Pi (q_1,q_2)=|q_1,q_2\rangle\rangle\langle\langle q_1,q_2| \equiv
|q_1\rangle\langle q_1 | \otimes |q_2\rangle\langle q_2|$ 
with $q_j \in \{x_j,p_j\}$, $j=1,2$ and $|q_j\rangle$ being quadrature
eigenstates. 
In order to implement a 
discrimination strategy, we should define an inference rule 
connecting  each possible outcome of the measurement to one of 
the two hypothesis: $H_{\tl}$, the noise is due to local interaction 
or $H_\tc$, the noise is due to common interaction with the environment.
If we denote by ${\mathcal D}_c \subset {\mathbb R}^2$ 
the region of outcomes leading to $H_\tc$, i.e. to infer a common noise,  
then the two-value POVM describing the overall
discrimination strategy is given by $E_\tc+E_\tl={\mathbb I}$, where
\begin{align}
E_\tc = \int\!\!\!\int_{\mathcal D_c}\!\! dq_1 dq_2\, \Pi(q_1,q_2) 
\qquad E_{\tl} = {\mathbb I} - E_\tc\,.
\end{align}
The success probabilities, i.e. those of inferring the
correct kind of noise are given by $P_j = \hbox{Tr}[\rho_j\,E_j]$, 
$j=L,C$ respectively, whereas the error probability, i.e. the
probability of chosing the wrong hypothesis is given by 
\begin{align}
p_\tq &= \frac12 \big(\hbox{Tr}[\rho_\tl\,E_\tc]+
\hbox{Tr}[\rho_\tc\,E_\tl]\big)
\nonumber \\ 
&= \frac12 \left(
1-  
\int\!\!\!\int_{\mathcal D_c}\!\! dq_1 dq_2\, \hbox{Tr}\left[
\Pi(q_1,q_2)\,(\rho_\tc-\rho_\tl)\right]
\right)
\end{align}
distribution of $(q_1,q_2)$.  
\par
The smaller is $P_\tq$, the more effective is the discrimination 
strategy. In order to suitably choose $\mathcal D_c$ we have analyzed
the behavior of the quantity
$$
p_\tq(q_1,q_2) = \hbox{Tr}\left[
\Pi(q_1,q_2)\,(\rho_\tc-\rho_\tl)\right], 
$$
in the $(q_1,q_2)$ plane. In Fig. \ref{wigner} we show a 
contourplot of $p_\tq(x_1,x_2)$ for a given input STS. This 
probability is squeezed along
the $x_1 x_2$ direction, since a common environment induces the
build-up of correlations between the quadratures. For this reason, we
choose ${\mathcal D}_c$ as the region between two straight lines
at $45^\circ$ and denote by $T$ its half-width. The same
argument holds also for SV and standard SV.
\par
In the top panels of Fig. \ref{fig1} we show a comparison between the
error probability of the four POVMs described above on
some particular STSs (left), standard form SVs (center) and SVs (right).
As it is apparent from the plot, the POVMs $\Pi (x_1,p_2)$ and 
$\Pi (p_1,x_2)$ are useless. In fact, the common
environment does not correlate these couples of quadratures.  On the
other hand, the POVM $\Pi (x_1,x_2)$, represented by the blue lines,
always outperforms $\Pi (p_1, p_2)$.  In the lower panels, we show the
optimal values of the half-width $T$ of the region $\mathcal D_c$ as 
a function of the interaction time for
the very same states. 
\section{Random input Gaussian states}
\label{s:random}
In this section, we address the optimization of the discrimination
protocol using Gaussian states as input and the optimal homodyne-based
POVM $\Pi (x_1,x_2)$. The main purpose 
is to figure out which Gaussian state leads to the optimal
discrimination protocol and understand which lab-friendly 
states, among the classes of STSs, SVs and SSVs, are 
the most performant ones.  We also analyze whether the 
efficiency of the discrimination protocol is
affected by some relevant properties of the input states. 
To this aim we evaluate the error probability $P_Q$ as a 
function of energy and entanglement, at fixed purity. 
We recall that the energy $E$ and purity $\mu$ of a zero-mean valued 
two-mode Gaussian state with covariance matrix $\bsigma$
are given by
\begin{align}
E(\bsigma) &= \mbox{Tr}\Big(\frac{\bsigma}{2} \Big)-1 \nonumber \\
\mu (\bsigma)& = \frac{1}{4 \sqrt{\mbox{det} \bsigma}},
\end{align}
while the entanglement may be easily determined on the base of the
PPT-criterion and quantified by the logarithmic
negativity, which is given by $$
\mathcal N = \mbox{max} \{0, -\log (2 \tilde{d}_1) \}
$$
where $\tilde{d}_1$ indicates the smallest symplectic eigenvalue
of the partially transposed covariance matrix. In this paper,
we prefer to directly use $\tilde{d}_1$
as a quantifier for entanglement: when $\tilde{d}_1< 1/2$,
the state is entangled, otherwise it is not.
\par
\begin{figure}[t!]
\includegraphics[width= \columnwidth]{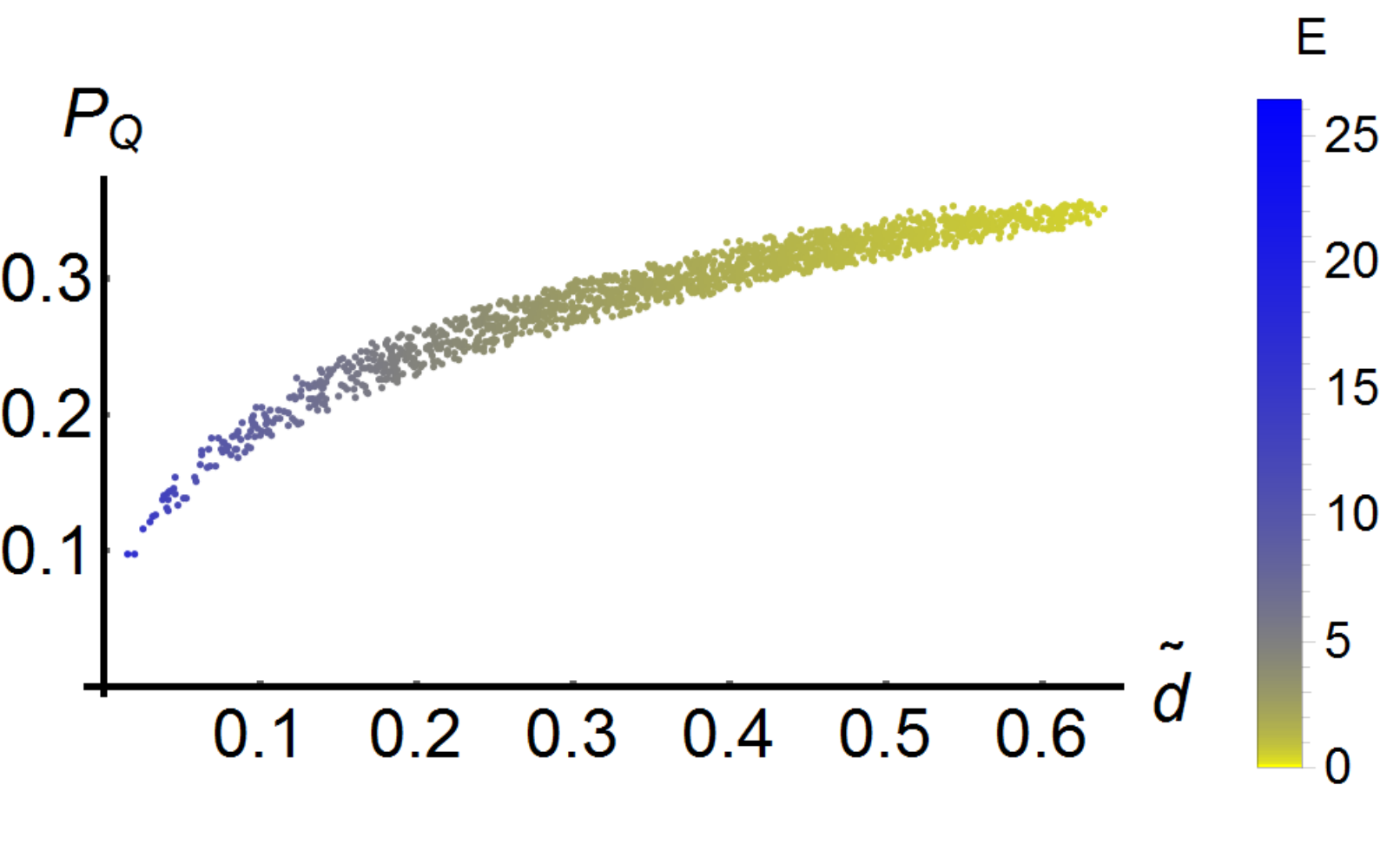}
\caption{\label{f:rand} (Color Online) Error probability $P_Q$ for
random Gaussian input states as a function of the smallest symplectic
eigenvalue $\tilde{d}_1$ with POVM $\Pi (x_1, x_2)$. The color scale
classifies the initial energy of the state. The error probability scales
with the entanglement and the energy of the input state.  We set
$\lambda_1 = \lambda_2 = \lambda =1, t=1$.} 
\end{figure}
\par
In Fig. \ref{f:rand}, we report the error probability of randomly
generated Gaussian states in standard form with purity $\mu =0.6$ at
fixed time $t=1$ as a function of the symplectic eigenvalue
$\tilde{d}_1$ of the input state, while the color scale classifies its
initial energy.  As is apparent from the figures, for non-unitary
purity, generating an always more entangled input state does not
necessarily imply an improvement  in the efficiency of the
discrimination protocol. The same happens with energy: the error
probability does not scale monotonicly with the energy stored in the
input state. Nevertheless, if we increase  the energy and the
entanglement of the input state at the same time, the error probability
lowers monotonicly.
\par
\begin{figure}[h!]
\includegraphics[width=  \columnwidth]{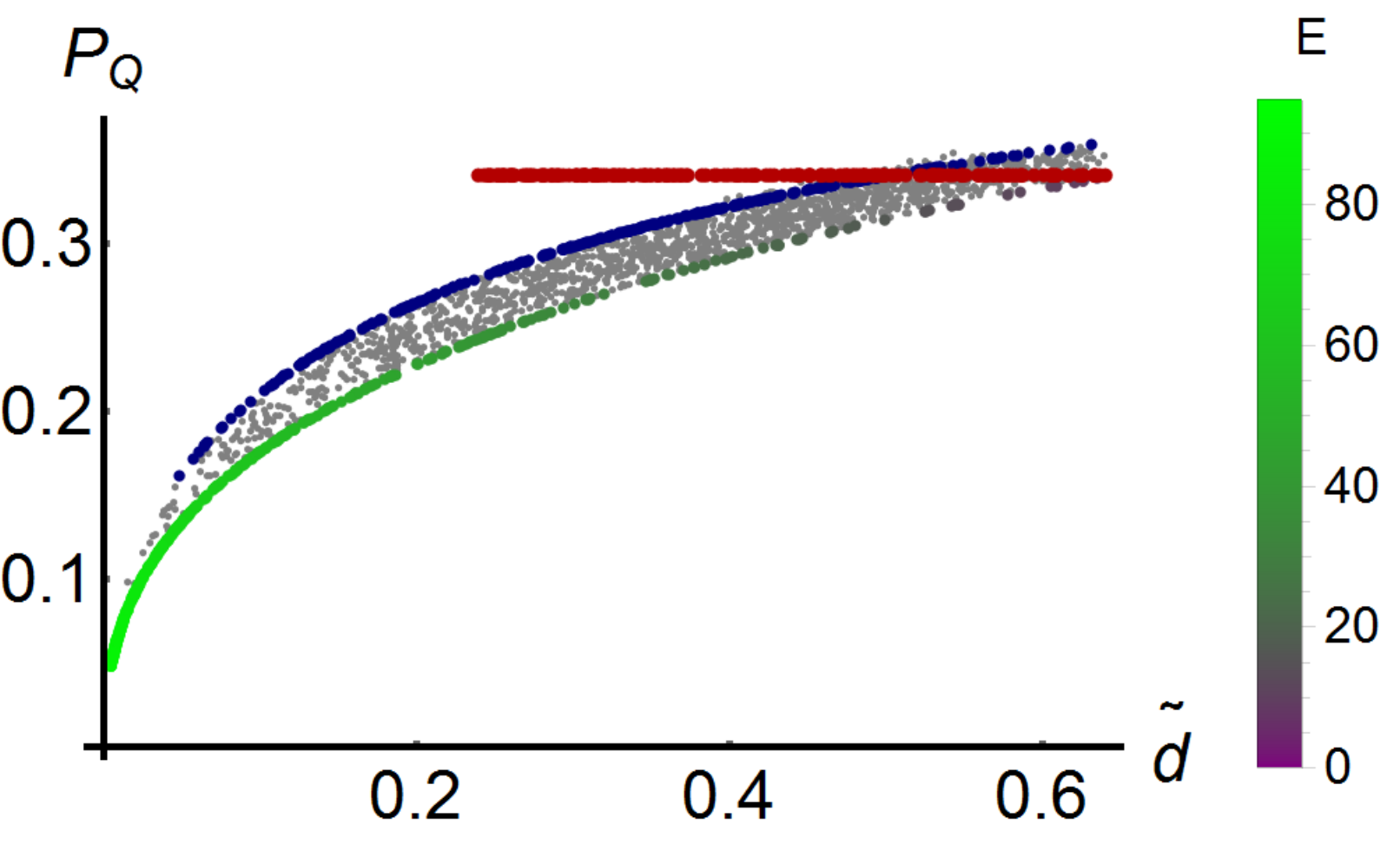}
\caption{\label{f:comparison} (Color Online) Error probability $P_Q$ as
a function of the smallest symplectic eigenvalue $\tilde d_1$ with POVM
$\Pi(x_1,x_2)$ for standard form random Gaussian states (gray dots),
STSs (blue upper curve), SVs (red straight line) and SSVs (lower line).
The color scale classifies the energies of the SSV states. The error
probability of SV states does not depend on the entanglement of the
input state. The most performant states are the SSVs.  We set $\lambda
=1$. } 
\end{figure}
\par
In Fig. \ref{f:comparison} we show how efficient STSs, SVs and SSVs are
with respect to all possible Gaussian states with the same purity ($\mu
=0.6$).  As a result, the most performant states are the SSVs: these
states form a lower bound for every random-generated state, so
representing the topmost suitable class state for discrimination
protocols.  One might make a conjecture that for SSVs entanglement might
be the only resource to discrimination: unfortunately, this is true as
long as purity is fixed, as the energy of SSVs monotonicly increases
with entanglement, but false in general. Concerning SVs and STSs, it is
worth noting that STSs are easily outperformed by any other standard
form Gaussian state and that the error probability achieved with input
SV states is not affected by a change in the initial entanglement (we
want to remember that SVs' covariance matrix is not in standard form,
this explains why the red curve steps over the region of the
standard form Gaussian states).

\begin{figure}[h!]
\includegraphics[width=\columnwidth]{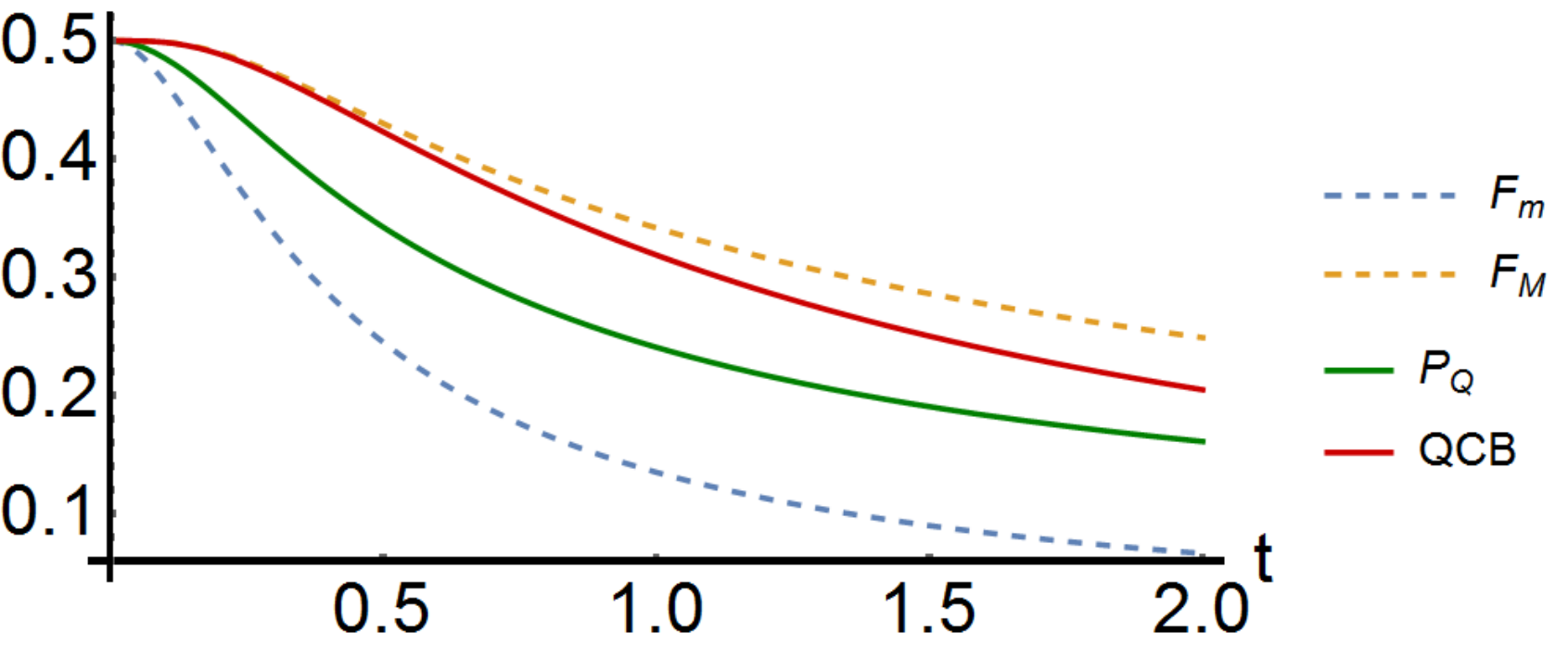}
\includegraphics[width=\columnwidth]{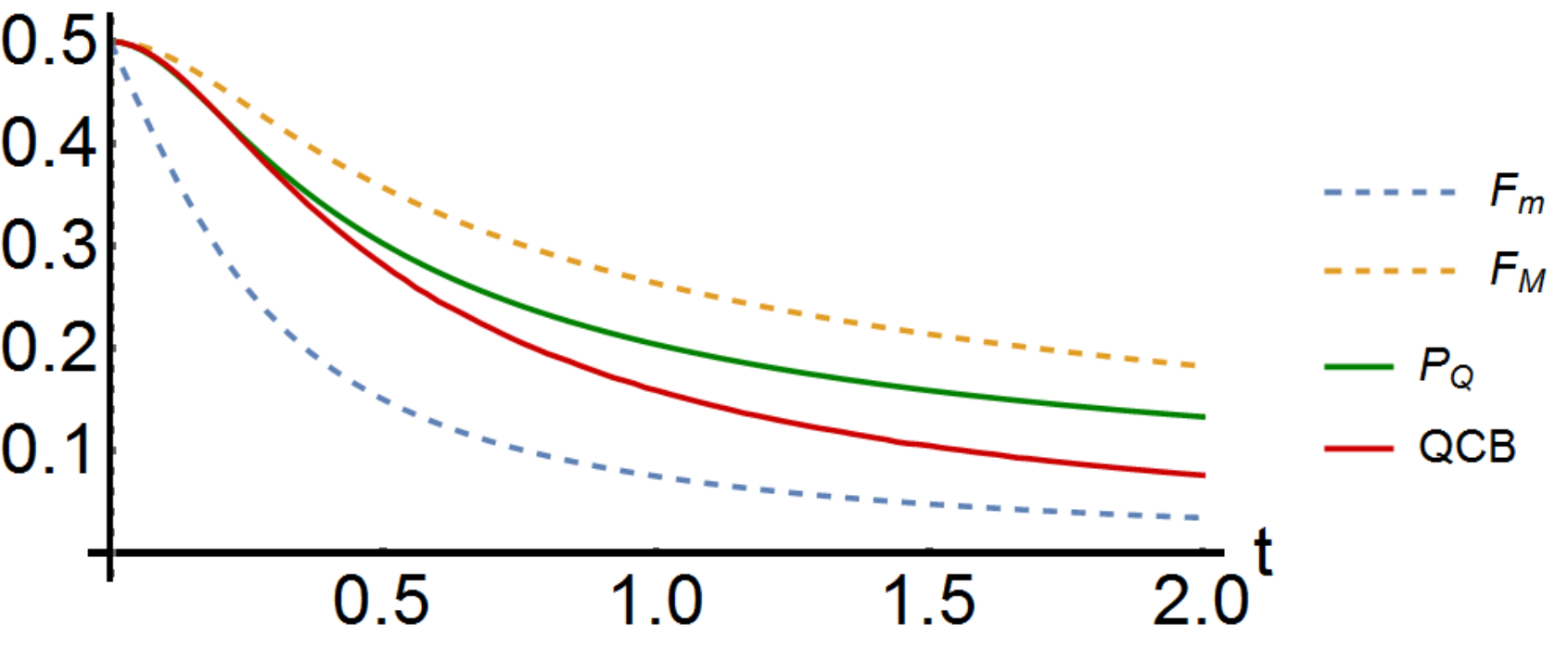}
\caption{\label{f:last} (Color Online) Comparison between error
probability and fidelity and QCB bounds for STS state (upper panel) and
SSV state (lower panel).  In both panels, the dashed lines represent the
upper bound $\mathcal F_M$ (orange line) and lower bound $\mathcal F_m$
(blue line), the green line represents the error probability
$P_Q$, the red line represents the QCB. Upper panel: we set
$\epsilon =1.956, \gamma = 0.6593, \lambda =1.0$. Lower panel: we set
$n=0.3333, r=1.470, \lambda =1$.} \end{figure}

Finally, in Fig. \ref{f:last} we compare the error probability achieved
by some lab-friendly states with the bounds we introduced in sec
\ref{sec:disc}.  In particular, we choose some highly performant
identically entangled STS and SSV.  The upper panel shows a comparison
between the error probability for a STS with the fidelity and the
Quantum Chernoff Bound. The double homodyne measurement yields an error
probability (green line) that beats the Quantum Chernoff Bound (red
line),
The lower panel shows a similar comparison for a SV state. In this case,
even though the SV state yields a lower error probability than a STS
does, the QCB can only be saturated
in the early dynamics.
\section{Conclusions}
\label{s:out}
In conclusion, we have addressed the design of effective strategies
to discriminate between the presence of local or common noise effect 
for a system made of two harmonic quantum oscillators
interacting with classical stochastic fields. 
The commoin noise scenario corresponds to the interaction 
of the two-mode quantum system with a common classical field, 
whereas the local one is 
described by coupling the oscillators to 
independent classical fields.
\par
We have shown that a discrimination protocol based on joint
homodyne detection of the position operators yields an error probability
that may outperform the Quantum Chernoff Bound, leading to 
a probability of error close to the Helstrom bound. In particular, we 
have shown that the QCB can be overtaken by means Gaussian states 
feasible with current technology. 
\par
Finally, we have shown that the error probability achieved with joint
homodyne measurement strictly depends on the properties of the input
state, as it lowers monotonicly with the energy and the entanglement of
the input state.
\begin{acknowledgments}
This work has been supported by UniMI through the H2020 Transition 
Grant 15-6-3008000-625, and by EU through the collaborative project 
QuProCS (Grant Agreement 641277). 
\end{acknowledgments}


\begin{thebibliography}{99}
\bibitem{wei01} U. Weiss {\em Quantum Dissipative 
Systems} (World Scientific, Singapore, 1999).
\bibitem{bre01} H. P. Breuer, F. Petruccione {\em 
The Theory of Open Quantum Systems} 
(Oxford University Press, Oxford, 2003).
\bibitem{har01} S. Haroche and J.-M. Raimond 
{\em Exploring the Quantum} 
(Oxford University Press, Oxford, 2006).

\bibitem{zur01} W. H. Zurek,"Decoherence and the transition from quantum
to classical" Phys. Today {\bf 44}, 36 (1991).

\bibitem{zur02} J. P. Paz, S. Habib, and W. H. Zurek, "Reduction of the
wave packet: Preferred observable and decoherence time scale", Phys. 
Rev. D {\bf 47}, 488 (1993).

\bibitem{ser1} M. G. A. Paris, A. Serafini, F. Illuminati, S. De Siena, 
"Purity of Gaussian states: measurement schemes
and time--evolution in noisy channels", 
Phys. Rev. A {\bf 68}, 012314 (2003). 

\bibitem{ser2} A. Serafini, S. De Siena, F. Illuminati, and M. G. A. Paris, 
"Minimum~decoherence~cat-like~states in~Gaussian~noisy~channels", 
J. Opt. B. {\bf 6}, S591 (2004). 

\bibitem{ser23}
A. Serafini, F. Illuminati, M. G. A. Paris, S. De Siena, 
"Entanglement and purity of two--mode Gaussian states in noisy
channels", Phys. Rev A {\bf 69}, 022318 (2004). 

\bibitem{ebe10} T. Yu, J. H. Eberly, 
"Entanglement Evolution in a Non-Markovian Environment"
Opt. Comm. {\bf 283}, 676 (2010).

\bibitem{bra01} D. Braun, "Creation of Entanglement by Interaction
with a Common Heat Bath", Phys. Rev. Lett. {\bf{89}} 277901
(2002).

\bibitem{zha01} Y. Zhao, G.H. Chen, "Two oscillators in a dissipative
bath", Physica A {\bf 317}, 13 (2003).

\bibitem{flo01} F. Benatti, R. Floreanini, M. Piani, "Environment in-
duced entanglement in Markovian dissipative dynamics", 
Phys. Rev. Lett. {\bf 91}, 070402 (2003).

\bibitem{pra01} J. S. Prauzner-Bechcicki, "Two-mode squeezed vacuum
state coupled to the common thermal reservoir", J. Phys. A: Math. 
Gen. {\bf 37}, L173 (2004).

\bibitem{con01} L. D. Contreras-Pulido, R. Aguado, "Entanglement 
between charge qubits induced by a common dissipative environment", 
Phys. Rev. B {\bf 77}, 155420 (2008).

\bibitem{paz01} J. P. Paz, A. J. Roncaglia, "Dynamics of the 
Entanglement between Two Oscillators in the Same Environment", 
Phys. Rev. Lett. {\bf 100}, 220401 (2008).


\bibitem{bro01} M. Brownnutt, M.  Kumph, P. Rabl and R. Blatt, 
"Ion-trap measurements of electric-field noise near surfaces", 
Rev. Mod. Phys. {\bf 87}, 1419 (2015).

\bibitem{gro01} S. Groeblacher, A. Trubarov, N. Prigge, G. D. Cole, M. As-
pelmeyer, J. Eisert, "Observation of non-Markovian
micromechanical Brownian motion", Nature Comm. {\bf 6},
7606 (2015).

\bibitem{pal01}  G. M. Palma, K.-A. Suominen, A. K. Ekert 
"Quantum Computers and Dissipation", Proc. R. Soc. London A {\bf 452}, 
567 (1996).

\bibitem{riv01} A. Rivas, M. Muller, "Quantifying spatial 
correlations of general quantum dynamics", 
New J. Phys. {\bf 17}, 062001
(2015).

\bibitem{zan01} P. Zanardi, M. Rasetti, "Noiseless Quantum Codes"
Phys. Rev. Lett. {\bf{79}}, 3306 (1997).

\bibitem{dua01} L.-M. Duan, G.-C. Guo, "Preserving Coherence in
Quantum Computation by Pairing Quantum Bits", Phys.
Rev. Lett. {\bf{79}}, 1953 (1997).

\bibitem{kwi01} P. G. Kwiat, A. J. Berglund, J. B. Altepeter, A.
G. White, "Experimental verification of decoherence-free 
subspaces", Science {\bf 290}, 498 (2000).

\bibitem{dic01} R. H. Dicke, "Coherence in 
Spontaneous Radiation Processes", 
Phys. Rev.{\bf{93}}, 99 (1954).


\bibitem{bar01} J. T. Barreiro et al. "An open-system 
quantum simulator with trapped ions", Nature {\bf 470}, 486 (2011).

\bibitem{ver01} F. Verstraete, M. M. Wolf, J. I. Cirac, "Quantum 
computation and quantum-state engineering driven 
by dissipation", Nature Phys. {\bf 5}, 633  (2009).

\bibitem{gar01} C.W. Gardiner, {\it{Handbook 
of Stochastic Methods}}, (Springer, Berlin, 1983).
\bibitem{hel01} J. Helm and W. T. Strunz 
"Quantum decoherence of two qubits", 
Phys. Rev. A {\bf{80}}, 042108 (2009).
\bibitem{hel02} J. Helm, W. T. Strunz, S. Rietzler, 
and L. E. W\"urflinger "Characterization of decoherence 
from an environmental perspective", Phys. Rev. A {\bf{83}}, 042103 (2011).

\bibitem{cro01} D. Crow and R. Joynt "Classical simulation of quantum
dephasing and depolarizing noise", Phys. Rev. A {\bf{89}}, 042123
(2014).

\bibitem{wit01} W. M. Witzel, K. Young, and S. Das Sarma "Converting a
real quantum spin bath to an effective classical noise acting on a
central spin", Phys. Rev. B {\bf{90}}, 115431 (2014).

\bibitem{str02}W. T. Strunz, L. D\'iosi, and N. Gisin "Open System
Dynamics with Non-Markovian Quantum Trajectories", Phys. Rev. Lett.
{\bf{82}}, 1801 (1999).

\bibitem{sto01}J. T. Stockburger and H. Grabert "Exact c-Number
Representation of Non-Markovian Quantum Dissipation", Phys. Rev. Lett.
{\bf{88}}, 170407 (2002).

\bibitem{tur01} Q.A. Turchette, C. J. Hyatt, B.E. King, C. A. Sackett,
D. Kielpinski, W. M. Itano, C. Monroe and D. J. Wineland "Decoherence
and decay of motional quantum states of a trapped atom coupled to
engineered reservoirs", Phys. Rev. A {\bf{62}}, 053807 (2000).

\bibitem{ast01}  O. Astafiev, Yu. A. Pashkin, Y. Nakamura, T. Yamamoto,
and J. S. Tsai "Quantum Noise in the Josephson Charge Qubit", Phys. Rev.
Lett. {\bf{93}}, 267007 (2004).

\bibitem{gal01} Y. M. Galperin, B. L. Altshuler, J. Bergli, and D. V.
Shantsev "Non-Gaussian Low-Frequency Noise as a Source of Qubit
Decoherence", Phys. Rev. Lett. {\bf{96}}, 097009 (2006).

\bibitem{abe01} B. Abel and F. Marquardt "Decoherence by quantum
telegraph noise: A numerical evaluation", Phys. Rev. B {\bf{78}},
201302(R) (2008).

\bibitem{str01} T. Grotz, L. Heaney and W.T. Strunz "Quantum dynamics in
fluctuating traps: Master equation, decoherence, and heating", Phys.
Rev. A, {\bf{74}}, 022102 (2006).

\bibitem{tra01} J. Trapani, M.G.A. Paris, "Non-divisibility vs backflow
of information in understanding revivals of quantum correlations for
continuous-variable systems interacting with fluctuating environments",
Phys. Rev. A {\bf{93}}, 042119 (2016).

\bibitem{hel03} C. W. Helstrom, {\em Quantum Detection 
and Estimation Theory} (Academic, New York, 1976).


\bibitem{aud02} K. M. R. Audenaert, M. Nussbaum, A. Szkola, and F. Verstraete,
Commun. Math. Phys. {\bf{279}}, 251 (2008).

\bibitem{pir01} S. Pirandola and S. Lloyd, Phys. 
Rev. A {\bf{78}}, 012331 (2008).
\bibitem{cals01} J. Calsamiglia, R. Munoz-Tapia, L. Masanes, A. Acin, and
E. Bagan, Phys. Rev. A {\bf{77}}, 032311 (2008).

\bibitem{aud01} K. M. R. Audenaert, J. Calsamiglia, R. Munoz-Tapia, E. Bagan,
Ll. Masanes, A. Acin, F. Verstraete, Phys. Rev. Lett. {\bf{98}}, 160501
(2007).

\bibitem{witt01} C. Wittmann, U. L. Andersen, M. Takeoka, D. Sych, G.
Leuchs, "Discrimination of binary coherent states using a homodyne
detector and a photon number resolving detector", Phys. Rev. A
{\bf{81}}, 062338 (2010).

\bibitem{oliv01} S. Olivares, S. Cialdi, F. Castelli, M.G.A. Paris,
"Homodyne detection as a near-optimum receiver for phase-shift-keyed
binary communication in the presence of phase diffusion", Phys. Rev. A
{\bf 87}, 050303(R), (2013).

\bibitem{pir02} S. Pirandola, "Quantum Reading of a Classical Digital
Memory", Phys. Rev. Lett. {\bf 106} , 090504 (2011).


\end{thebibliography}
\end{document}